\documentclass{JHEP3}

\usepackage{amssymb}
\usepackage{amsmath}
\usepackage{graphicx}
\usepackage{latexsym}

\newcommand{\GeV}{\text{GeV}}
\newcommand{\BR}{\text{BR}}
\newcommand{\DM}{{\text{DM}}}

\title{CDMS II result and Light Higgs Boson Scenario of the MSSM}

\author{
Masaki Asano\\
Department of Physics, Tohoku University, Sendai 980-8578, Japan\\
E-mail: \email{masano@tuhep.phys.tohoku.ac.jp}
}
\author{
Shigeki Matsumoto\\
Department of Physics, University of Toyama, Toyama 930-8555, Japan\\
E-mail: \email{smatsu@sci.u-toyama.ac.jp}
}
\author{
Masato Senami
\thanks{
Now, the author is in Department of Micro Engineering,
Kyoto University, Kyoto 606-8501, Japan
}\\
ICRR, the university of Tokyo, Kashiwa, Chiba 277-8582, Japan\\
E-mail: \email{senami@me.kyoto-u.ac.jp}
}
\author{
Hiroaki Sugiyama\\
Department of Physics, Ritsumeikan University, Kusatsu, Shiga 525-8577, Japan\\
E-mail: \email{hiroaki@fc.ritsumei.ac.jp}
}

\abstract{
The CDMS Collaboration has reported two candidate events for dark matter. If the events are due to the elastic scattering of dark matter, the dark matter would be a WIMP dark matter with its mass of the order of 10-100~GeV and its scattering cross section with a nucleon is about 10$^{-44}$-10$^{-43}$~cm$^2$. We show that such a dark matter is properly realized as a neutralino dark matter in the light higgs boson scenario of the MSSM\@. The lightest higgs boson mass can be lighter than $114.4$~GeV in the scenario because of a suppressed interaction between higgs boson and $Z$ bosons. As a result, a large scattering cross section between the dark matter and ordinary matter is obtained.
}

\preprint{TU-865}

\begin{document}

\section{ Introduction }\label{sect:intro}

Existence of dark matter in our universe has already been established thanks to recent cosmological observations such as the Wilkinson Microwave Anisotropy Probe (WMAP) experiment~\cite{Komatsu:2008hk}, and now we know that about 23\% of the energy density of the universe is provided by dark matter. We know, however, very little of what the dark matter is. Many attempts to identify the dark matter have been performed so far~\cite{DM Review}. Very recently, two candidate events for dark matter have been reported by the Cryogenic Dark Matter Search (CDMS) collaboration~\cite{Ahmed:2009zw}. The signals may be due to the scattering between a nucleus in the detector and dark matter in the halo associated with our galaxy. Though the events are still consistent with background fluctuation at the probability of 23\%, the events may give us important information about the dark matter.

If the two events are caused by the elastic scattering of the dark matter inside the detector, the dark matter would be a Weakly Interacting Massive Particle (WIMP) with its mass of 10-100~GeV~\cite{Kopp:2009qt}. It is also important to notice that the scattering cross section between dark matter and a nucleon is of the order of 10$^{-43}$-10$^{-44}$~cm$^2$ if the scattering occurs through spin-independent interactions. Such a large cross section can be obtained from the process in which the higgs boson is exchanged, when the interaction between the higgs boson and two dark matters is large enough. This kind of scenario has already been discussed by many papers even in a very short period after the report of the CDMS collaboration~\cite{Previous activities}.

In this letter, we propose another scenario to account for the CDMS result, where the large scattering cross section is obtained by a light higgs boson with mass less than 114.4~GeV\@. One might think that such a light higgs boson is not favored by the LEP2 experiments~\cite{LEPSM}. However the higgs boson indeed can be light in the framework of the light higgs boson scenario (LHS) of the minimal supersymmetric standard model (MSSM)~\cite{LHS1,LHS2,Asano:2007gv,LEPSUSY}. Furthermore, as shown in Ref.~\cite{Asano:2007gv}, the scenario is consistent with the WMAP experiment as well as other experimental constraints such as $b \rightarrow s \gamma$ and $B_s \rightarrow \mu^+ \mu^-$.

The essential reason why the higgs boson can be lighter than the LEP bound ($m_h \geq 114.4$~GeV) in the LHS is as follows. There are two higgs doublet fields in the MSSM, which leads to two CP-even neutral higgs bosons. By choosing an appropriate mixing between these two, the coupling constant between the lightest higgs boson and $Z$ bosons can be significantly smaller than that of the standard model (SM). The mass of the lightest higgs boson therefore can be lighter than the LEP bound, because the bound comes from the process $e^+ e^- \rightarrow Zh$. In addition, the LHS can explain the 2.3~$\sigma$ level excess of the events at 98~GeV in the LEP2 experiments~\cite{LEPSM, LEPSUSY}, which is very difficult to be explained in the SM\@.

We consider the SUSY model with non-universal scalar masses for the higgs multiplets (NUHM), which is one of concrete models realizing the LHS being compatible with the grand unified theory (GUT). Details of our setup will be shown in the next section. In order to clarify the region consistent with many experimental constraints in a broad parameter space, we use the Markov Chain Monte Carlo (MCMC) method, which will be discussed in section~\ref{sect:MCMC}. Finally, in section~\ref{sect:DD}, it is found that the model predicts the bino-like neutralino dark matter with its mass 50-250~GeV\@. Interestingly, the scattering cross section between the dark matter and a nucleon can be around 10$^{-44}$~cm$^2$, which is nothing but the value requested by the CDMS result.

\section{ LHS of the MSSM }\label{sect:LHS}

With the assumption of CP conservation, two higgs doublet fields ($H_u, H_d$) in the MSSM leads to the spectrum of the higgs sector composed of three neutral and two charged scalar bosons: two CP-even higgs bosons ($h$ and $H$, where $h$ is lighter than $H$), CP-odd higgs boson ($A$), and charged higgs bosons ($H^\pm$). The mass eigenstates of the CP-even higgs bosons are given by the mixing states between neutral components of $H_u$ and $H_d$,
\begin{eqnarray}
  \begin{pmatrix}
    h \\ H
  \end{pmatrix}
  =
  \begin{pmatrix}
    -\sin\alpha & \cos\alpha \\ 
    \cos\alpha & \sin\alpha 
  \end{pmatrix}
  \begin{pmatrix}
    {\rm Re}~H_d^0 \\ 
    {\rm Re}~H_u^0
  \end{pmatrix},
\end{eqnarray}
where $\alpha$ is the angle introduced to diagonalize the mass squared matrix of CP-even higgs bosons,
\begin{eqnarray}
  \small
  \begin{pmatrix}
    m_A^2 s_\beta^2 + m_Z^2 c_\beta^2 + \Delta_{dd} &
    - (m_A^2 + m_Z^2) s_\beta c_\beta + \Delta_{du} \\
    - (m_A^2 + m_Z^2) s_\beta c_\beta + \Delta_{du} &
    m_A^2 c_\beta^2 + m_Z^2 s_\beta^2 + \Delta_{uu}
    \rule{0ex}{3ex} 
  \end{pmatrix}.
\end{eqnarray}
The mass of the CP-odd higgs boson ($Z$ boson) is denoted by $m_A$ ($m_Z$) and $c_\beta (s_\beta) \equiv \cos \beta (\sin \beta$), where the ratio of the vacuum expectation value of the higgs doublet fields is given by $\tan\beta = \langle H_u^0 \rangle/\langle H_d^0 \rangle$. The radiative correction to each component in the matrix is represented by $\Delta_{ii}$, where its detailed expression can be found in Ref.~\cite{BookDrees}.

The higgs boson has been searched using the process $e^+ e^- \to Z h$ in the LEP2 experiments. The coupling constant between $h$ and $Z$ bosons is given by $g_{ZZh} = g_{ZZh}^{\rm (SM)} \sin (\beta - \alpha)$ in the MSSM, where $g_{ZZh}^{\rm (SM)}$ is the corresponding coupling constant in the SM\@. As a result, the lightest higgs boson $h$ can be lighter than the LEP bound ($m_h \geq 114.4$~GeV) when the coupling constant $g_{ZZh}$ is significantly smaller than $g_{ZZh}^{\rm (SM)}$~\cite{LHS1,LHS2,Asano:2007gv,LEPSUSY}.

Since the angle $\beta$ is expected to be as large as $\pi/4 $-$ \pi/2$, a large mixing angle $\alpha$ is required to realize a small $g_{ZZh}$. Such a large angle is obtained when $m_A$ is small enough, which leads to the situation that all higgs bosons in the MSSM are at the scale of ${\cal O}(100)$ GeV. Moreover, suppressed $\sin (\beta - \alpha)$ leads to enhanced $\cos (\beta - \alpha)$, which results in a large coupling constant $g_{ZZH}$ between heavy higgs boson and $Z$ bosons. Therefore, the heavy higgs boson should satisfy the constraints of the LEP higgs boson search. In addition, the coupling constant $g_{ZAh}$ between the lightest higgs boson, CP-odd higgs boson, and $Z$ boson is also proportional to $\cos (\beta - \alpha)$. This coupling leads the $e^+ e^- \to h A$ process at the LEP2 experiments. Hence, the constraint for this process should also be taken into account. Fortunately, this constraint is not severe due to the P-wave suppression, because the coupling $g_{ZAh}$ originates in a derivative interaction.

The LHS cannot be realized in the constrained MSSM, which is widely used to study the MSSM\@. If the universal masses of sfermions and gauginos are fixed to derive $m_h < 114.4$ GeV, the masses of other SUSY particles becomes too small. In order to realize the LHS, the masses of the higgs doublet fields should be different from others as $m_{H_u(H_d)}^2 = (1+\delta_{H_u(H_d)}) m_0^2$. It is reasonable because the higgs multiplets are not necessarily in the same multiplet of GUT with other scalar particles. The simplest model with this boundary condition is the NUHM\@. Hence, we adopt this model as a reference model to investigate the LHS~\cite{LHSLFV}. The NUHM has six free parameters, $(m_0, m_{1/2}, A_0, \tan \beta, \mu, m_A)$. The first three parameters $(m_0, m_{1/2}, A_0)$ are universal sfermion mass, gaugino mass, and tri-linear coupling, which are defined at the GUT scale $M_G$. Other parameters, $\tan\beta$, higgsino mass $\mu$, and $m_A$ are defined at the electroweak scale. This parameterization allows us to treat the masses of two higgs doublet fields at $M_G$ as free parameters. Using these values at the $m_Z$ scale, a boundary condition at $M_G$ is derived by the renormalization group running. Masses of SUSY particles then run back from $M_G$ to $m_Z$. We use ISAJET 7.75~\cite{ISAJET} to evaluate the renormalization group running.

\section{ Markov Chain Monte Carlo }\label{sect:MCMC}

\begin{center}
  \begin{table*}[t]
    \begin{tabular}{|l|l|c|c|c|c|c|c|c|}
      \hline
      & Constraints & References \\
      \hline
      Relic abundance of dark matter ($\Omega_\DM h^2$) &
      $0.1099 \pm 0.0062$ &
      \cite{Komatsu:2008hk} (WMAP only)
      \\
      $\BR(b\to s\gamma)$ &
      $(3.52 \pm 0.25)\times 10^{-4}$ &
      \cite{Barberio:2008fa}
      \\
      $\rho$-parameter ($\Delta\rho$) &
      $0 \pm 0.0009$ &
      \cite{Amsler:2008zzb} (p.~137)
      \\
      $\BR(B_s\to \mu^+ \mu^-)$ &
      $< 5.8\times 10^{-8}$ &
      \cite{:2007kv}
      \\
      Upper bound on $g_{ZZh}$ ($g_{ZZH}$) &
      Function of $m_h$ ($m_H$) &
      \cite{LEPSUSY}
      \\
      Upper bound on $g_{ZAh}$ ($g_{ZAH}$) &
      Function of $m_h (m_H) + m_A$ &
      \cite{LEPSUSY}
      \\
      Lightest neutralino mass &
      $> 50.3\,\GeV$ &
      \cite{LEPSUSYweb}
      \\
      Chargino mass &
      $> 103.5\,\GeV$ &
      \cite{LEPSUSYweb}
      \\
      Right-handed selectron mass &
      $> 99.9\,\GeV$ &
      \cite{LEPSUSYweb}
      \\
      Right-handed smuon mass &
      $> 94.9\,\GeV$ &
      \cite{LEPSUSYweb}
      \\
      Right-handed stau mass &
      $> 86.6\,\GeV$ &
      \cite{LEPSUSYweb}
      \\
      Sneutrino masses &
      $> 94\,\GeV$ &
      \cite{Abdallah:2003xe}
      \\
      Stop and sbottom masses &
      $> 95\,\GeV$ &
      \cite{LEPSUSYweb}
      \\
      Gluino mass &
      $> 308\,\GeV$ &
      \cite{:2007ww}
      \\
      Squark masses (1st and 2nd generations) &
      $> 379\,\GeV$ &
      \cite{:2007ww}
      \\
      \hline
    \end{tabular}
    \caption{\small Constraints used in our Markov Chain Monte Carlo analysis.}
    \label{tab: constraints}
  \end{table*}
\end{center}

Markov Chain Monte Carlo method is a random sampling algorithm, which constructs a series of parameter sets (Markov chain)~\cite{recipes}. The samples of the chain obey a distribution which is proportional to a given distribution function. The distribution function of our interest is a posterior probability distribution function of model parameters $x$ under experimental data $D$. Bayes' theorem tells us that the posterior probability distribution function $P(x|D)$ satisfies the following equation,
\begin{eqnarray}
  P(x|D) = \frac{ P(D|x) P(x) }{ \sum_{x^\prime} P(D|x^\prime) P(x^\prime) },
\end{eqnarray}
where $P(x)$ is the prior probability function reflecting our knowledge about the model parameters $x$, while $P(D|x)$ is representing the likelihood for the distribution function of experimental data $D$ at given model parameters $x$. In our analysis, a linearly flat prior has been used for $P(x)$, where $P(x)dx$ gives a constant probability.

\begin{figure}[t]
  \begin{center}
    \includegraphics[origin=c, angle=0,width=7.2cm]{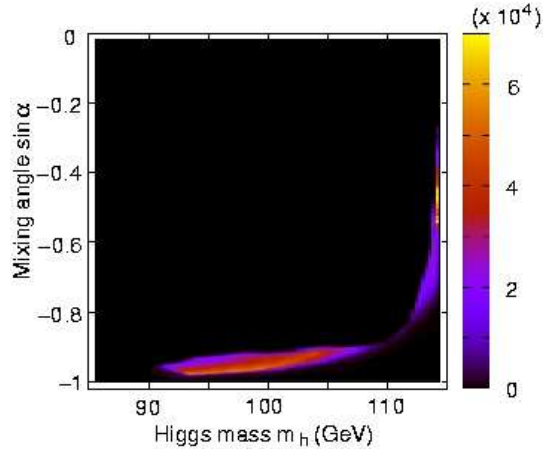}
    \caption{The distribution of samples. The yellow (bright) region is of a large number of samples.}
    \label{fig:mh-alpha}
  \end{center}
\end{figure}

For experimental data $D$, we have used constraints shown in Table~\ref{tab: constraints}. The branching ratio $\BR(b\to s\gamma)$ has been calculated with SusyBSG~1.3.1~\cite{Degrassi:2007kj}. Other observables and parameters are given by  micrOMEGAs~2.2.CPC~\cite{Belanger:2008sj}. We then use a Gaussian distribution for constraints on $\Omega_\DM h^2$, $\BR(b \to s\gamma)$, and $\Delta\rho$, while other constraints have been introduced simply as boundaries of the model parameter space. In addition, since we are interested in the LHS, we concentrate on our exploration in the range $m_h < 114.4$~GeV\@. Finally, we have generated a chain of about $2.3\times 10^7$ samples. In figure~\ref{fig:mh-alpha}, the distribution of the samples is shown on the ($m_h$, $\sin\alpha$) plane. Two distinct regions can be seen in the figure. In the region with low $|\alpha|$, the property of $h$ is essentially the same as the one in previous studies for $m_h > 114$~GeV\@. On the other hand, the high $|\alpha|$ region is nothing but the region discussed in section~\ref{sect:LHS}, which is characteristic of the LHS\@. We thus use only the samples in the high $|\alpha|$ region by imposing a condition $m_h < 110$~GeV in the following analysis,
which is about $2.0\times 10^7$ samples.

\section{ Direct Detection of Dark Matter}\label{sect:DD}

\begin{figure}[t]
  \begin{center}
    \includegraphics[origin=c, angle=0,width=7cm]{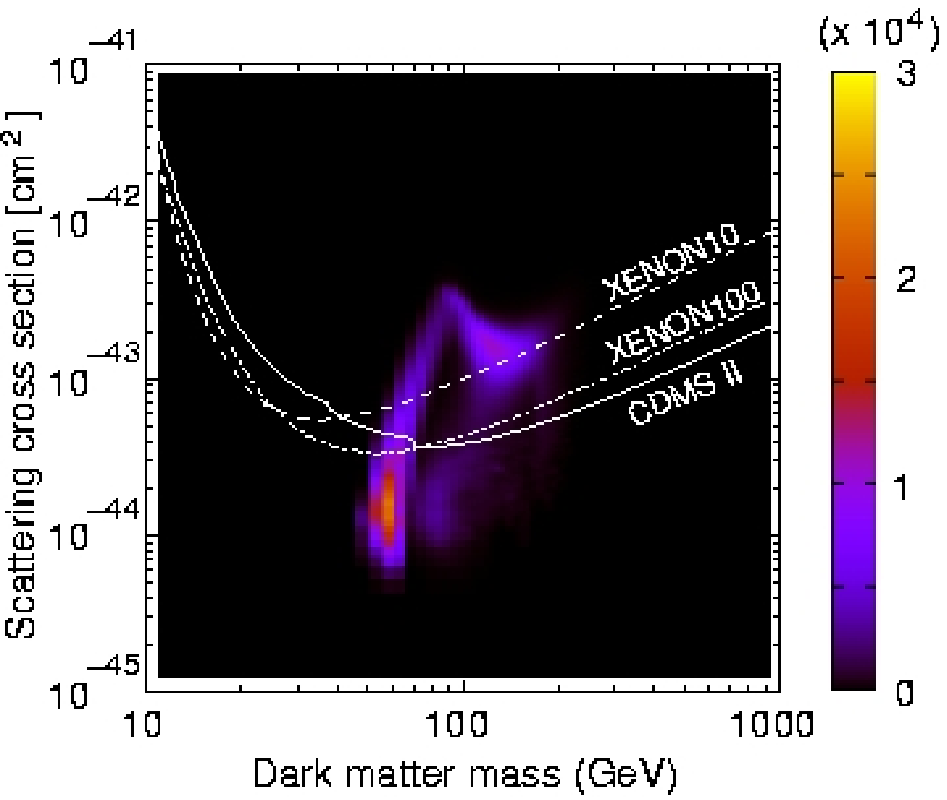}
    \includegraphics[origin=c, angle=0,width=7cm]{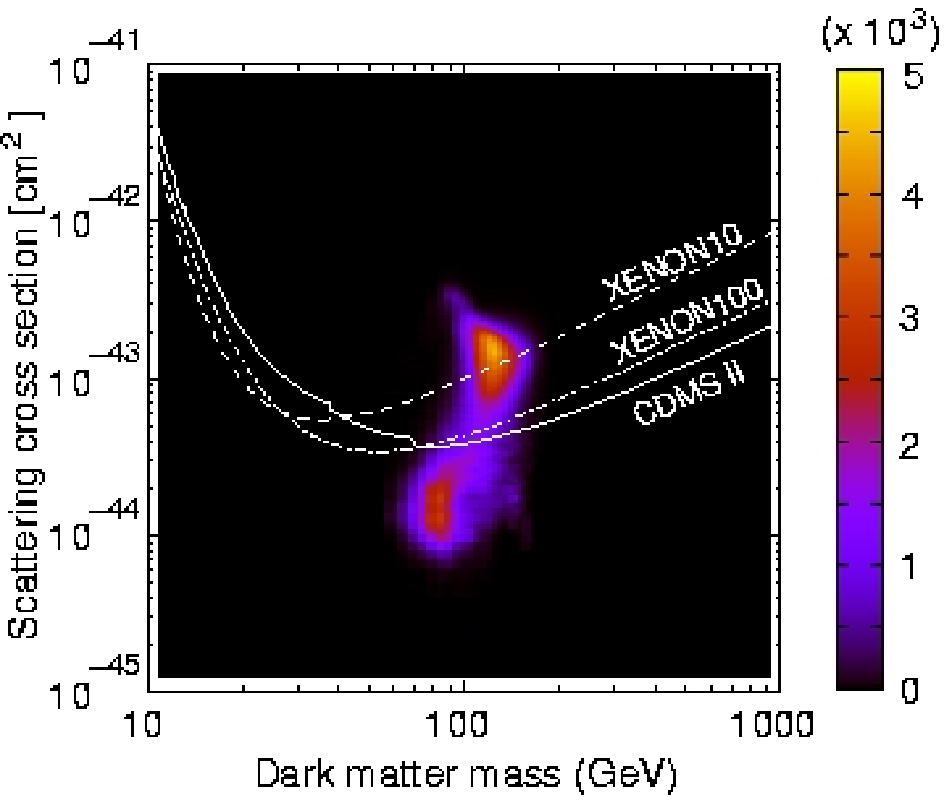}
    \caption{
      The prediction for direct detection experiments
     of the dark matter in the LHS\@.
      The figures show the distribution of samples
     which satisfy $m_h < 110\,\GeV$.
      The right figure is given by samples
     which satisfy an additional condition on $g-2$ of muon.
      The yellow (bright) region is of a large number of samples.
}
    \label{fig:DD}
  \end{center}
\end{figure}
\begin{figure}[t]
  \begin{center}
    \includegraphics[origin=c, angle=0,width=7cm]{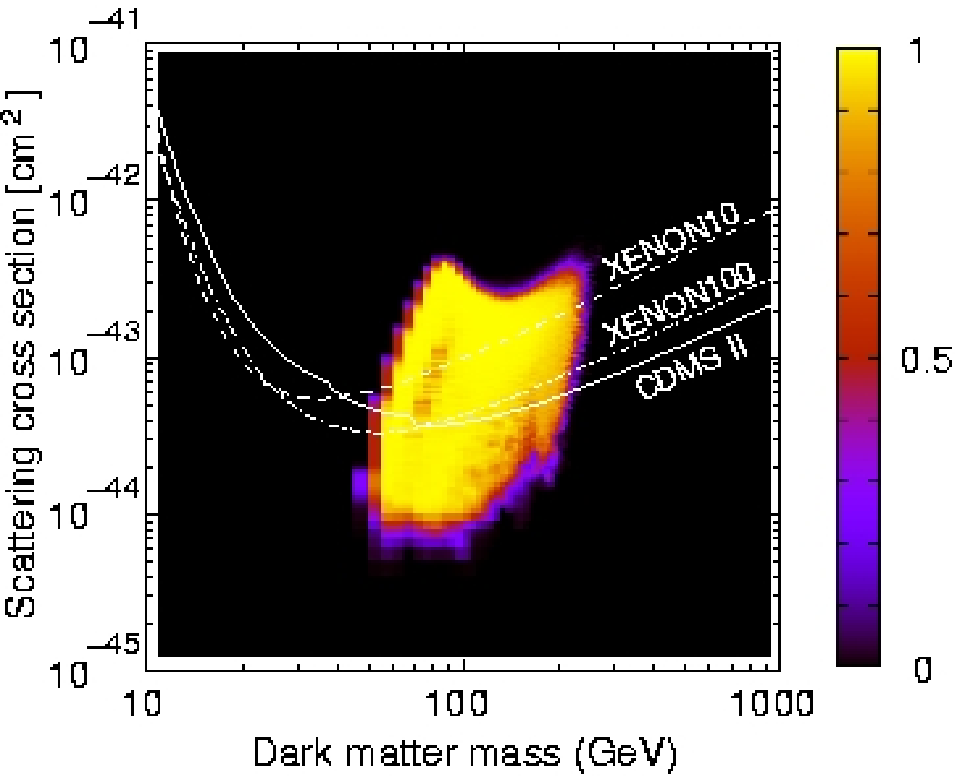}
    \includegraphics[origin=c, angle=0,width=7cm]{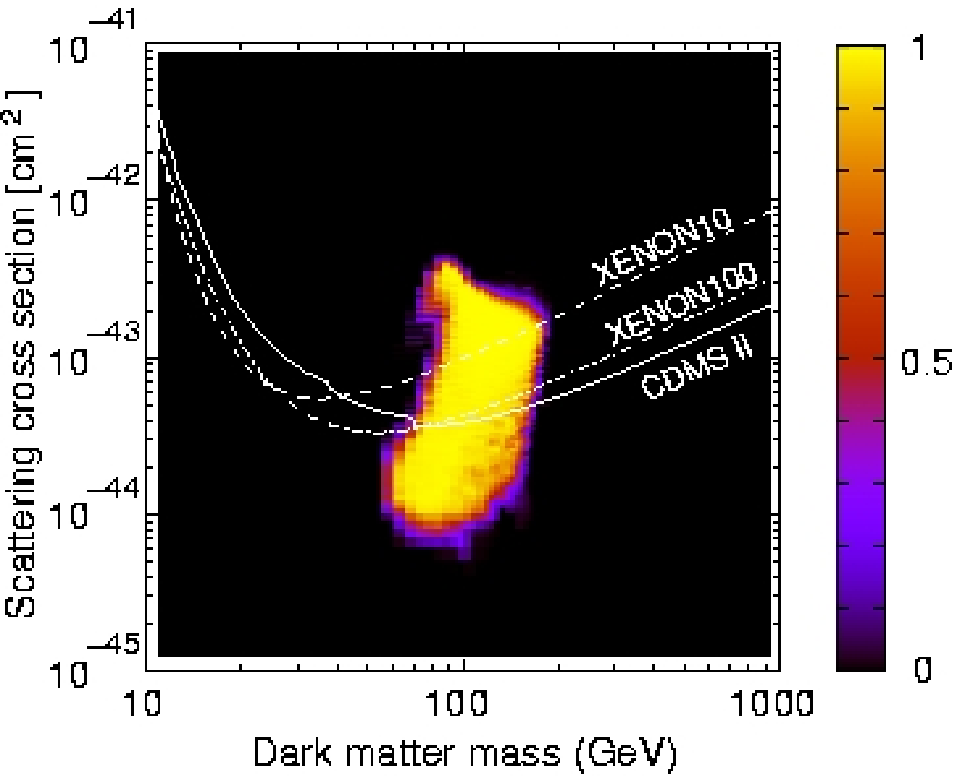}
    \caption{
      Likelihood maximum which is obtained by calculating the maximum
     in each cell with samples which satisfy $m_h < 110\,\GeV$.
      The right figure is given by samples
     which satisfy an additional condition on $g-2$ of muon.
      The yellow (bright) region is of a large value of
     likelihood maximum in a cell,
     where constraints on $\Omega_\DM h^2$, $\BR(b\to s\gamma)$,
     and $\Delta\rho$ can be satisfied in a good accuracy.
 }
    \label{fig:DD_lhmax}
  \end{center}
\end{figure}

 Since all higgs bosons in the LHS are at the electroweak scale,
detection rates at direct detection experiments of dark matter,
which is the lightest neutralino in our analysis,
are expected to be large~\cite{LHS1, Asano:2007gv,DirectLHS}.
 The distribution of the MCMC samples is shown in figure~\ref{fig:DD}
with cells which have the same size in the linear scale.
 The right panel is the result with samples
which satisfy also a condition
$\Delta a_\mu = ( 29.2 \pm 2\times 8.6 )\times 10^{-10}$%
~\cite{Amsler:2008zzb}~(p.~482)
for the anomalous magnetic moment ($g-2$) of muon,
while the condition is not used in the left one.
 The horizontal axis of these figures is the mass of the dark matter,
and the vertical one is the spin-independent cross section
between the dark matter and a nucleon.
 In our calculation,
the $y$ parameter, which characterizes the strangeness component in a nucleon,
is set to be zero~\cite{Ohki:2008ff}.
 Current bounds on the scattering cross section
from XENON10~\cite{Angle:2007uj}, XENON100~\cite{Aprile:2010um},
and CDMS~II~\cite{Ahmed:2009zw,Ahmed:2008eu} experiments
are also shown in these figures as dashed, dot-dashed, and solid lines, respectively.
 Note that regions of a large number of samples in figure~\ref{fig:DD}
do not mean the regions of good agreement with experimental results
because the number of constraints
(i.e.\ $\Omega_\DM h^2$, $\BR(b \to s\gamma)$, and $\Delta\rho$)
is less than the number (six) of parameters in the NUHM\@.
 The cell with many samples simply shows
that the values of the dark matter mass and
the scattering cross section in the cell are easily obtained by the MCMC method
because these values are predicted in a wide area of the parameter space of the NUHM\@.
 In figure~\ref{fig:DD_lhmax},
we also plot likelihood maximum%
\footnote
{
 We define the likelihood so that
it is proportional to the product
of Gaussian distribution functions
for $\Omega_\DM h^2$, $\BR(b \to s\gamma)$, and $\Delta\rho$.
 The set of the central values for them in Table~\ref{tab: constraints}
makes the likelihood unity.
}
which is obtained by calculating the maximum in each cell.
 The right panel in figure~\ref{fig:DD_lhmax}
is obtained by samples which satisfy
a condition for $g-2$ of muon.
 A large value of the likelihood means that
constraints in Table~\ref{tab: constraints} are well satisfied.
 It is seen that
most of the region of the samples agrees with the constraints
in a good accuracy.
 The cross section is predicted to be around $10^{-44}$-$10^{-43}~\text{cm}^2$
in most of samples with the mass of the dark matter being around 100~GeV\@.
 If the CDMS~II events are really coming from
the elastic scattering of dark matter,
the LHS can easily explain those events.
By comparing the left and right panels in figures~\ref{fig:DD} and \ref{fig:DD_lhmax},
it can be seen that a large part of the region is consistent
with the result of $g-2$ of muon.

 The lower bound on the mass of the dark matter
in figures~\ref{fig:DD} and \ref{fig:DD_lhmax} are controlled by the bound on
the lightest neutralino in Table~\ref{tab: constraints},
which is originally determined by those of charginos
through the GUT relation on gaugino masses.
 The relic abundance of the dark matter around the lower bound is
governed by the s-channel diagram to $b \bar{b}$,
in which the CP-odd higgs boson mediates.
 Such a region vanishes in right panels of figures~\ref{fig:DD} and \ref{fig:DD_lhmax}.
 On the other hand, the upper bound on the mass of the dark matter
comes again from those of charginos.
 Too heavy charginos are not favored because the cancellation
between the contributions from the charginos
and light charged higgs bosons in the LHS is required to be consistent
with the constraint on the $b\rightarrow s\gamma$ process.
 In the region 
where the dark matter mass is larger than about 100~GeV,
the relic abundance of dark matter is mostly governed
by the processes whose final states are composed of higgs bosons~\cite{Kim:2008uh}
or the coannilation process between the LSP and stau;
 for samples of large scattering cross section around $10^{-43}\,\text{cm}^2$,
the former processes are dominant,
while small scattering cross section below around $10^{-44}\,\text{cm}^2$
is realized by the coannihilation process.
 The coannihilation process is efficient
in the samples used for
the right panels of figures~\ref{fig:DD} and \ref{fig:DD_lhmax}.

 Even if the two events reported by the CDMS collaboration are due to background fluctuation, the neutralino dark matter in the LHS will be tested in near future at SuperCDMS, XENON100, and XMASS experiments~\cite{FutureDD} because the region predicted by the scenario is not far away from the current bounds.

\section*{ Acknowledgments }
This work is supported in part by the Grant-in-Aid for Science Research, Ministry of Education, Culture, Sports, Science and Technology, Japan (No.16081211 for S.~M.) and the Grant-in-Aid for the Global COE Program Weaving Science Web beyond Particle-matter Hierarchy from the Ministry of Education, Culture, Sports, Science and Technology of Japan (M.~A.).

\end{document}